\documentclass[12pt]{iopart}
\usepackage{iopams}
\usepackage{graphicx}
\usepackage[compress]{cite}
\usepackage{graphicx}
\usepackage{bm}
\usepackage{mathptmx}
\usepackage{amssymb}
\usepackage{latexsym}

\begin{document}

\title [Investigation of spectral properties of $^{11}$Be]{Investigation of spectral properties of $^{11}$Be in breakup reactions}

\author{D S Valiolda$^{1, 2, 3}$, D M Janseitov$^{1, 2, 3}$ and V S Melezhik$^{3, 4}$}

\address{$^1$ Al-Farabi Kazakh National University, Almaty, 050074, Republic of Kazakhstan}
\address{$^2$ Institute of Nuclear Physics, Almaty, 050032, Republic of Kazakhstan}
\address{$^3$ Bogoliubov Laboratory of Theoretical Physics, Joint Institute for Nuclear Research, Dubna, 141980, Russian Federation}
\address{$^4$ Dubna State University, Dubna, 141980, Russian Federation}

\ead{janseit.daniar@gmail.com}
\vspace{10pt}

\begin{abstract}
We investigate the breakup of the $^{11}$Be halo nuclei on a light target ($^{12}$C) within quantum-quasiclassical approach in a wide range of beam energy (5–67 MeV/nucleon) including bound states and low-lying resonances in different partial and spin states of $^{11}$Be. The obtained results are in good agreement with existing experimental data at 67 MeV/nucleon.  We also demonstrate that the developed computational scheme can be used for investigation of nuclei spectral properties in low-energy breakup experiments on different targets.

\end{abstract}

%
\vspace{2pc}
\noindent{\it Keywords}: halo nucleus, breakup cross section, low-lying resonances, computational scheme
%
%
\maketitle
%
%

\section{Introduction}

Breakup reactions, in which the valence neutron is removed from the projectile by its interaction with a target nucleus, have played a valuable role in probing the exotic nuclear structures, such as halo nuclei~\cite{tan1996, hansen1987, nakamura1994}. During the breakup reactions of these nuclei, the angular distributions of neutrons measured in coincidence with the core nuclei~\cite{anne1990} are strongly forward peaked and the parallel momentum distributions of the core fragment have very narrow widths~\cite{orr1995, bertul1992, banerj1995}, which provide a confirmation of their halo structure. $^{11}$Be is a famous example of one-neutron halo nuclei, where the loosely bound valence neutron has a large spatial extension with respect to the corresponding core.

The breakup of the halo nuclei has been investigated theoretically by several authors using a number of different approaches: the time-dependent (TD) models~\cite{kido1994, melezhik1999, melezhik2003,  kido1996, esbensen1995, fallot2002}, the coupled-channel technique with a discretized continuum (CDCC)~\cite{tostevin2001, yahiro1982}, models based on the eikonal approximation ~\cite{glaub1959}, particularly the dynamical eikonal approximation (DEA)~\cite{baye2005, gold2006}. Each model has its own peculiarities. In particular, an advantage of the non-perturbative time-dependent model~\cite{melezhik1999, melezhik2003} is the inclusion in the calculation of all higher-order effects in the relative motion of the breakup fragments, which provides a fully dynamical description of the projectile excitation caused by both the Coulomb and the nuclear interactions between the projectile and the target.

In our previous works~\cite{valiolda2022, valipepan2022}, we have carried out with the extended quantum-quasiclassical model based on the non-perturbative time-dependent approach~\cite{melezhik1999, melezhik2003} a comprehensive analysis of the influence of resonant states on the breakup of a $^{11}$Be halo nucleus on a heavy target ($^{208}$Pb) in a wide energy region (5 - 70 MeV/nucleon). It was shown that including of the low-lying resonances ($5/2^{+}$, $3/2^{-}$ and $3/2^{+}$) of $^{10}$Be+n system made a significant contribution to the breakup cross section, which gives a better agreement with existing experimental data~\cite{fukuda2004}.

Here, we perform calculations of the breakup cross section of $^{11}$Be on a light carbon target at beam energies 5 - 67 MeV/nucleon with the curvilinear trajectory of the projectile and including low-lying resonances ($5/2^{+}$, $3/2^{-}$, $3/2^{+}$) of $^{10}$Be+n system. The obtained results are in good agreement with existing experimental data at 67 MeV/nucleon~\cite{fukuda2004}. In the low energy region there are only a few theoretical results~\cite{hebborn2019, hebborn2018}, with which we compare our calculations. The performed analysis demonstrates the possibility of studying spectral properties of halo nuclei in their breakup reactions.

This paper is organized as follows. After the brief introduction of the subject with
literature survey, in Sect. 2, we introduce a description of the non-perturbative time-dependent model with linear trajectories and a quantum-quasiclassical approach with “real” trajectories of $^{11}$Be in breakup reactions. In the third section the convergence of the computational scheme is demonstrated and the obtained results are presented. In the fourth section the investigation of spectral structure of $^{11}$Be in breakup reactions is given. The last section is devoted to concluding remarks.

\section{Description of the model}
\subsection{Non-perturbative time-dependent approach with linear trajectories of $^{11}$Be}

This approach is based on the integration of the time-dependent three-dimensional Schr\"odinger equation for halo neutron for describing its dynamics during collision of $^{11}$Be with a target~\cite{melezhik1999, melezhik2003}.

The halo neutron is treated as a structureless particle weakly bound by the potential \textit{U(\textbf{r})} to the $^{10}$Be core nucleus, where \textit{\textbf{r}} is the the relative variable between the neutron and the core. The time-dynamics of the halo neutron relative the $^{10}$Be core in the breakup reaction $^{11}$Be+$^{12}$C $\rightarrow ^{10}$Be+n+$^{12}$C is depicted by the time-dependent Schr\"odinger equation

\begin{equation}
 \label{Eq1}
i\hslash\frac{\partial}{\partial\textit{t}}\Psi(\textbf{r},t)=H(\textbf{r},t)\Psi(\textbf{r},t)=[H_{0}(r)+V(\textbf{r},\textbf{R},t)]\Psi(\textbf{r},t)
\end{equation}

\noindent in the projectile rest frame, where $\Psi(\textbf{r},t)$ is the wave packet of the neutron relative the $^{10}$Be core. In this expression

\begin{equation}
H_{0}(\textbf{r})=-\frac{\hslash^{2}}{2\mu}\Delta_{r}+U(\textbf{r})
\label{one}
\end{equation}
is the Hamiltonian describing the halo nucleon-core system with reduced mass $\mu=m_{n}m_{c}/M$, where $m_{n}, m_{c}$ and M=$m_{n}+m_{c}$ are the neutron, $^{10}$Be-core, and $^{11}$Be masses, respectively. The potential $U$(\textbf{r}) consists of the sum of the \textit{l}-dependent Woods-Saxon potential \textit{$V_{l}(r)$} and the spin-orbit interaction \textit{$V_{l}^{s}(r)(\textbf{l$\cdot$s}) = -V_{l}^{s}\frac{1}{r}\frac{d}{dr} f(r)(\textbf{l$\cdot$s})$}, where $f(r)=1/(1+\exp(\frac{r-R_{0}}{a}))$. The parameters of potential have a standard value: $R_{0}$=2.585 fm, $a$= 0.6 fm and the depth $V_{l}^{s}$= 21 MeV $fm^{2}$.

The parameters of the spherical Woods-Saxon potentials $V_{l}(r)=-V_{l}/(1+\exp(\frac{r-R_{0}}{a}))$, describing the energy spectrum of $^{11}$Be nucleus, had been determined as $V_{l}$= 62.52 MeV (\textit{l}-even) and $V_{l}$= 39.74 MeV ($l$-odd)~\cite{baye2005} in order to reproduce the $1/2^{+}$ ground state of $^{11}$Be at -0.503 MeV, the $1/2^{-}$ excited state at -0.183 MeV and two resonance states $5/2^{+}$ and $3/2^{+}$ with the position of peaks at $E_{5/2^{+}}$= 1.232 MeV and $E_{3/2^{+}}$= 3.367 MeV~\cite{NNDC, ershov2014}. For these states, the radius is $R_{0}$= 2.585 fm and the diffuseness is \textit{a}= 0.6 fm. To fix the position of the $3/2^{-}$ resonance ($l$=1) close to the experimental~\cite{NNDC} and theoretical~\cite{ershov2014} value $E_{3/2^{+}}$= 2.789 MeV we tuned the set of parameters as $V_{l}$= 6.8 MeV, $R_{0}$= 2.5 fm and \textit{a}= 0.35 fm in our recent works~\cite{valiolda2022, valipepan2022}. For $l\geq3$, the spherical potential $V_{l}(r)$ was set to zero. More details of parameterization of potentials between the neutron and $^{10}$Be core and how the resonant states were included in the analysis of the breakup reaction were discussed in (see table 1 of ~\cite{valiolda2022}), \cite{valejpfm2022}.

The time-dependent potential $V(\textbf{r},\textbf{R},t)$ in Eq.(1) simulates the interaction of the target with the projectile. It was assumed to be purely Coulombic for breakup reactions with a heavy target ($^{208}$Pb) at collision energies around 70 MeV/nucleon~\cite{melezhik1999, melezhik2003, valiolda2022},  which is defined as $V_{C}(\textbf{r},\textbf{R},t)=\frac{Z_{c}Z_{t}e^{2}}{|m_{n}\textbf{r}/M+\textbf{R}(t)|}-\frac{Z_{c}Z_{t}e^{2}}{R(t)}$, where $Z_{c}$ and $Z_{t}$ are charge numbers of the core and target, respectively, and $\textbf{R}(t)$ is the relative coordinate between the projectile and the target. As it has been shown in previous studies with time-dependent non-perturbative approach~\cite{valiolda2022, valipepan2022}, the contribution of the nuclear part of the projectile-target interaction in the breakup cross sections on a heavy target is significant for lower beam energies (30 - 5 MeV/nucleon). In this work we evaluate this effect in the case of light target ($^{12}$C) using the approach of optical potential $\Delta V_{N}(\textbf{r},\textbf{R},t)=V_{cT}(\textbf{r}_{cT}(t))+V_{nT}(\textbf{r}_{nT}(t))$ for the nuclear interaction between the target and projectile:

\begin{equation}
V(\textbf{r},\textbf{R},t)=V_{C}(\textbf{r},\textbf{R},t)+\Delta V_{N}(\textbf{r},\textbf{R},t)\,.
\label{one}
\end{equation}
Here, $\textbf{r}_{cT}(t)$ and $\textbf{r}_{nT}(t)$ are the core-target $\textbf{r}_{cT}(t)=\textbf{R}(t)+m_{n}\textbf{r}/M$ and neutron-target $\textbf{r}_{nT}(t)=\textbf{R}(t)-m_{c}\textbf{r}/M$ relative variables and optical potentials $V_{cT}$  and $V_{nT}$ have the form:

\begin{equation}
V_{xT}(r_{xT})=-V_{x}f_{x}(r_{xT}, R_{R}, a_{R})-iW_{x}f_{x}(r_{xT}, R_{I}, a_{I})\,
\label{one}
\end{equation}
with Woods-Saxon form-factors $f_{x}(r_{xT},R_{R},a_{R}) = 1/(1+\exp(\frac{r_{xT}-R}{a}))$, where $x$ stands for either core or neutron.

The analytical expressions of such potentials were obtained by selecting the parameters of general form-factors so as to fit the calculated scattering cross sections onto experimental data. A compilation of optical potentials for different projectiles and targets can be found in Ref.~\cite{hebborn2019, parey1976, bechetti1969}.  We use here the parameters of the optical potentials (4) from the work~\cite{capel2004}, which are given in Table I. The potential of the core-target interaction is proposed by Al-Khalili, Tostevin, and Brooke~\cite{khalili1997} consistent with the elastic scattering of $^{10}$Be on $^{12}$C (denoted as ATB in the following). For the n- $^{12}$C interaction, commonly used parametrization of Becchetti and Greenlees~\cite{bechetti1969} (BG) is considered. The expediency of using this parameterization of the nuclear part of the interaction was analyzed in the work~\cite{bechetti1969}.

\begin{table}[h!]
\caption{\label{jlab1}Parameters of the core-target~\cite{khalili1997} and neutron-target~\cite{bechetti1969} optical potentials at 67 MeV/nucleon.}
\label{tab:2}
\begin{tabular}{lllllll}
\br
c or n & $ V_{x} (MeV) $ & $ W_{x} (MeV) $ & $ R_{R} (fm)$ & $ R_{I} (fm)$ &
$ \textit{a}_{R} (fm) $ & $ \textit{a}_{I} (fm) $ \\
\noalign{\smallskip}\hline\noalign{\smallskip}
 $^{10}$Be & 123.0 & 65.0 & 3.33  & 3.47  & 0.80 & 0.80  \\
n & 34.54 & 13.40 & 2.68 & 2.88 & 0.75 & 0.58  \\
\br
\end{tabular}
\end{table}

In early works on non-perturbative time-dependent approach~\cite{melezhik1999, melezhik2003} the straight-line trajectories $\textbf{R}(t)=\textbf{b}+\textbf{v}_{0}t$ were used, where $\textbf{b}$ is the impact parameter orthogonal to the initial velocity of the projectile  $\textbf{v}_{0}$. However, advancement into the region of low collision energies, where the deviation of the projectile trajectory from a straight line increases, required to go in beyond this approximation~\cite{valiolda2022}.

\subsection{Quantum-quasiclassical approach with “real” trajectories of $^{11}$Be}

In our previous work~\cite{valiolda2022}, we have extended the description of the $^{11}$Be breakup on $^{208}$Pb to low energies using the quantum-quasiclassical approach developed in~\cite{melezhik2004, melezhik2021, melezhik2001, melezhik2023j} and successfully applied previously in various problems of atomic physics. This made it possible to calculate with satisfactory accuracy the cross section of this reaction at low collision energies of up to 5 MeV/nucleon. Here we apply this approach to describe the breakup of $^{11}$Be on a light target $^{12}$C. In this regards, we have to note that the idea of using more realistic projectile trajectory is not new one. Thus, classical Coulomb trajectories were used in the work of Fallot et. al on breakup of $^{11}$Be~\cite{fallot2002} and in the work of Melezhik and Baye on breakup of $^{17}$F~\cite{melezhik2001}.

In this approach, simultaneously with the time-dependent Schr\"odinger equation (1) for the halo-nucleon wave function $\Psi(\textbf{r},t)$ we integrate the set of Hamilton equations

\begin{equation}
\label{Eq10}
\frac{d}{dt}\textbf{P} = -\frac{\partial}{\partial \textbf{R}}H_{BP}(\textbf{P}, \textbf{R}, t) \,,\,\frac{d}{dt}\textbf{R} = \frac{\partial}{\partial \textbf{P}}H_{BP}(\textbf{P}, \textbf{R}, t)
\end{equation}
describing relative projectile-target dynamics. Here, the classical Hamiltonian $H_{BP}(\textbf{P}, \textbf{R}, t)$ is given by

\begin{equation}
\label{Eq11}
\fl H_{BP}(\textbf{P}, \textbf{R}, t) = \frac{\textbf{P}^2}{2M} + \langle\Psi(\textbf{r},t)| V_{C}(\textbf{r}, \textbf{R}, t)|\Psi(\textbf{r},t)\rangle \simeq \frac{\textbf{P}^2}{2M}-\frac{Z_{C}Z_{T}e^{2} m_{n}}{M} \langle\Psi(\textbf{r},t)|\frac{(\textbf{r} \cdot \textbf{R})}{R^{3}}|\Psi(\textbf{r},t)\rangle
\end{equation}
where the term $\langle\Psi(\textbf{r},t)|V_{C}(\textbf{r}, \textbf{R}, t)|\Psi(\textbf{r},t)\rangle$ represents the quantum-mechanical average of the projectile-target interaction over the halo-nucleon density instantaneous distribution $|\Psi(\textbf{r},t)|^2$ during the collision. Constructing the effective classical Hamiltonian $H_{BP}$ which supposed to be real one, we have neglected the optical potential (4) having imaginary part and acting for impact parameters below $b\sim$ 10 fm, in the region where the breakup cross section drops sharply with decreasing $b$~\cite{melezhik2003, valiolda2022}. Thus, the Hamiltonian (6) defined in such a way has a parametric dependence on the halo-neutron position $\textbf{R}(t)$ at every time moment. This quantum-quasiclassical model permits to include in the computational scheme the deformation of the projectile trajectories and energy transfer between the target and projectile and vice versa during collisions. The required stability and accuracy of the integration of Eq. (5) simultaneously with the time-dependent Schr\"odinger equation (1) with the same step of integration over time was ensured by using a computational finite-difference scheme developed in~\cite{melezhik2021} based on the Störmer–Verlet method~\cite{hairer2006}.

The total breakup cross section is calculated as a function of the relative energy $E$ between the emitted neutron and the core nucleus including neutron interaction with the core in the final state of the breakup process by the formula~\cite{melezhik1999, melezhik2003, valiolda2022, melezhik2001}:

\begin{equation}
\frac{d\sigma_{bu}(E)}{dE}=\frac{4\mu\textit{k}}{\hslash^{2}}\int_{b_{min}}^{b_{max}}\sum_{j=l+s}\sum_{lm}|\int\phi_{ljm}(k,r)Y_{lm}(\hat{r})\Psi(\mathbf{r},T_{out})d\mathbf{r}|^{2}bdb\,.
\label{one}
\end{equation}
Here $\Psi(\mathbf{r},T_{out})$ is the neutron wave packet at the end of the collision process $^{11}$Be+$^{12}$C $\rightarrow ^{10}$Be+n+$^{12}$C (at $t=T_{out}$), which is found by numerical integration of the hybrid system of equations consisting of the time-dependent Schr\"odinger equation (1) and the classical Hamilton equations (5). $\phi_{ljm}(k,r)$ is the radial part of the eigenfunction of the Hamiltonian $H_{0}(\textbf{r})$ (2) in the continuum spectrum ( $E=k^2\hbar^2/(2\mu)>0$), normalized to a spherical Bessel function $\textit{j}_{l}(kr)$ as  $kr\rightarrow\infty$ if $U$(\textbf{r})=0, $Y_{lm}(\hat{r})$ are the spherical harmonics. Summation over ($l, m$) in (7) includes all 16 partial waves up to $l_{max}$= 3 inclusive, as in~\cite{valiolda2022}.

The time evolution starts at initial time $T_{in}$ and stops at final time $T_{out}$ by iteration over $N_{T}$ time steps $\bigtriangleup\textit{t}$ as explained in~\cite{melezhik1999, melezhik2003}. The initial (final) time $T_{in} (T_{out})$ has to be sufficiently big $\vert\textit{T}_{in}\vert, T_{out}\rightarrow+\infty$, fixed from the demand for the time-dependent potential $V(\textbf{r},\textbf{R},t)$ to be negligible at the beginning (end) of the time evolution at $t$=$T_{in}$($T_{out}$). Following the investigation performed in~\cite{valejpfm2023}, the time interval is fixed as  $T_{in}$= -10 $\hslash$/MeV and $T_{out}$=10 $\hslash$/MeV, the time step $\bigtriangleup\textit{t}$ equals to 0.02 $\hslash$/MeV.
Technically, the most difficult part when numerically integrating the system of equations (1), (5) is finding $\Psi(\mathbf{r}, t)$ - solving the three-dimensional time-dependent Schr\"odinger equation (1). To solve it, a two-dimensional discrete-variable representation (2D DVR) is used to approximate the desired wave function $\Psi (r,\Omega,t)$ in terms of angular variables $\Omega=(\theta,\varphi)$~\cite{melezhik1999, melezhik2003, melezhik1997}. Moreover, as shown in our previous work~\cite{valiolda2022}, as the collision energy of $^{11}$Be with a heavy target decreases, the convergence of this approach in the number of basis functions  $N$ of the 2D DVR (which is equal to the number of grid points in angular variables $N=N_{\theta}\times N_{\varphi}$) slows down. In the next section, we investigate the convergence of the 2D DVR in $N$ for our problem of the breakup of $^{11}$Be on a light target in the entire energy range under study.

For discretizing with respect to the radial variable \textit{r}, a sixth-order (seven point) finite-difference approximation on a quasiuniform grid has been used on the interval $r\in[0,r_{m}]$ with $r_{m}$= 600 fm. The grid has been realized by the mapping $r\rightarrow$\emph{x} of the initial interval onto $x\in[0,1]$ by the formula $r=r_{m}(e^{8x}-1)/((e^{8}-1))$~\cite{valiolda2022, melezhik1997}. The choice of the edges of integration over radial variables and convergence of the method with respect to radial meshes was discussed in~\cite{valejpfm2023}.

In the calculation of the breakup cross section the choice of edges of integration over impact parameters $b_{min}$ and $b_{max}$ must be carefully tested. For breakup reaction $^{11}$Be+$^{12}$C $\rightarrow ^{10}$Be+n+$^{12}$C the evolution is computed from impact parameters $b_{min}$= 0 fm up to $b_{max}$. The step $\vartriangle{b}$ is chosen in order to ensure the convergence of the integral in Eq. (7). It varies from $\bigtriangleup\textit{b}$= 0.25 fm at small b up to $\bigtriangleup\textit{b}$= 2 fm at large b as in~\cite{capel2004}. The inclusion of a strong interaction between the target and the projectile using the optical potential leads to a faster convergence of the integral (7) along the upper limit $b_{max}$ (as it was shown in~\cite{valejpfm2023} at $b_{max}$= 150 fm).
More details of numerical integration can be found in works~\cite{melezhik1999, melezhik2003, valiolda2022}. 

\section{Results and discussions} 

In our recent work~\cite{valejpfm2023}, the convergence with respect to $N$ of the time-dependent approach with linear trajectories of projectile for the breakup of the $^{11}$Be nucleus on a light target ($^{12}$C) was investigated at 67 MeV/nucleon and also accuracy of the numerical technique was discussed there. Here, we investigate the convergence in $N$ of the quantum-quasiclassical approach  including the curvature of the trajectory of the $^{11}$Be in collision with $^{12}$C-target (effect important at low collision energies of $^{11}$Be on $^{208}$Pb-target)~\cite{valiolda2022}. The main task of our investigation is to extend a quantum-quasiclassical approach for calculation of the breakup cross sections at low beam energies.

Therefore, the breakup cross sections $d\sigma_{bu}(E)/dE$ were computed with this approach on the 2D DVR basis functions extended from $N$= 49 ($N_{\theta}=N_{\varphi}=$7) to $N$= 225 ($N_{\theta}=N_{\varphi}=$15) with including two bound states (ground $1/2^{+}$ and first excited $1/2^{-}$ states) and three low-lying resonances ($5/2^{+}$, $3/2^{-}$, $3/2^{+}$) of $^{11}$Be at beam energy of E= 20 MeV/nucleon. Fig. 1 shows that it is sufficient to use ($N_{\theta}=N_{\varphi}=$15) angular grid points $N$= 225 (2D DVR basis functions) for computing the breakup cross section of $^{11}$Be on a $^{12}$C with demanded accuracy of the order of a few percents. This number of 2D DVR basis functions was used in the subsequent calculations.

\begin{figure}[htp]
\centerline{\includegraphics[width=11cm]{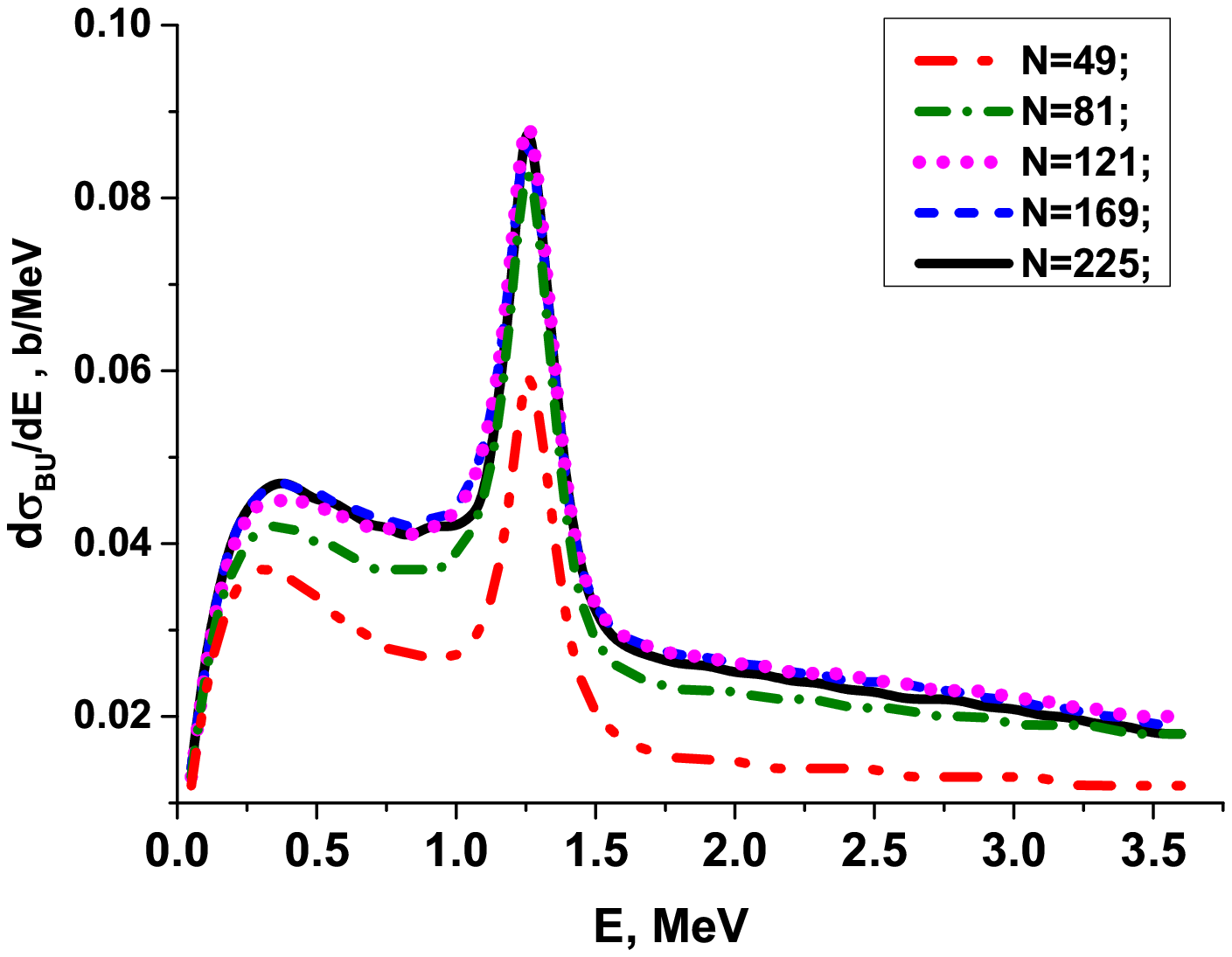}}
\caption{The convergence of the calculated with quantum-quasiclassical approach (1), (5) breakup cross section $d\sigma_{bu}(E)/dE$ (7) of $^{11}$Be at 20 MeV/nucleon as a function of the number $N=N_{\theta}\times\textit{N}_{\varphi}$ of angular grid points (2D DVR basis functions).} 
\label{fig:3}
\end{figure}

\begin{figure}[hbt]
\begin{minipage}[h]{0.5\linewidth}
\center{\includegraphics[width=1\linewidth]{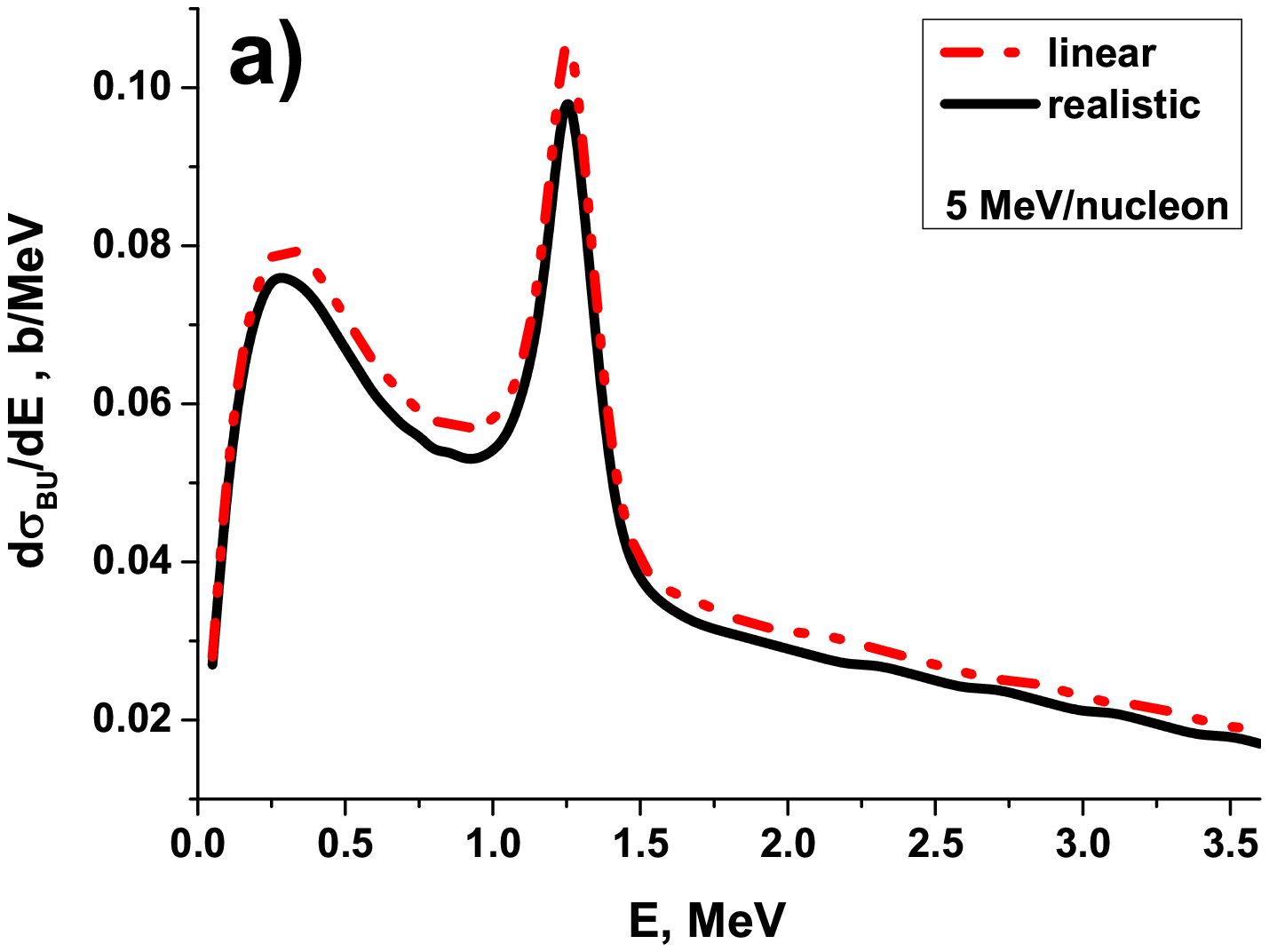}}
\end{minipage}
\hfill
\begin{minipage}[h]{0.49\linewidth}
\center{\includegraphics[width=1\linewidth]{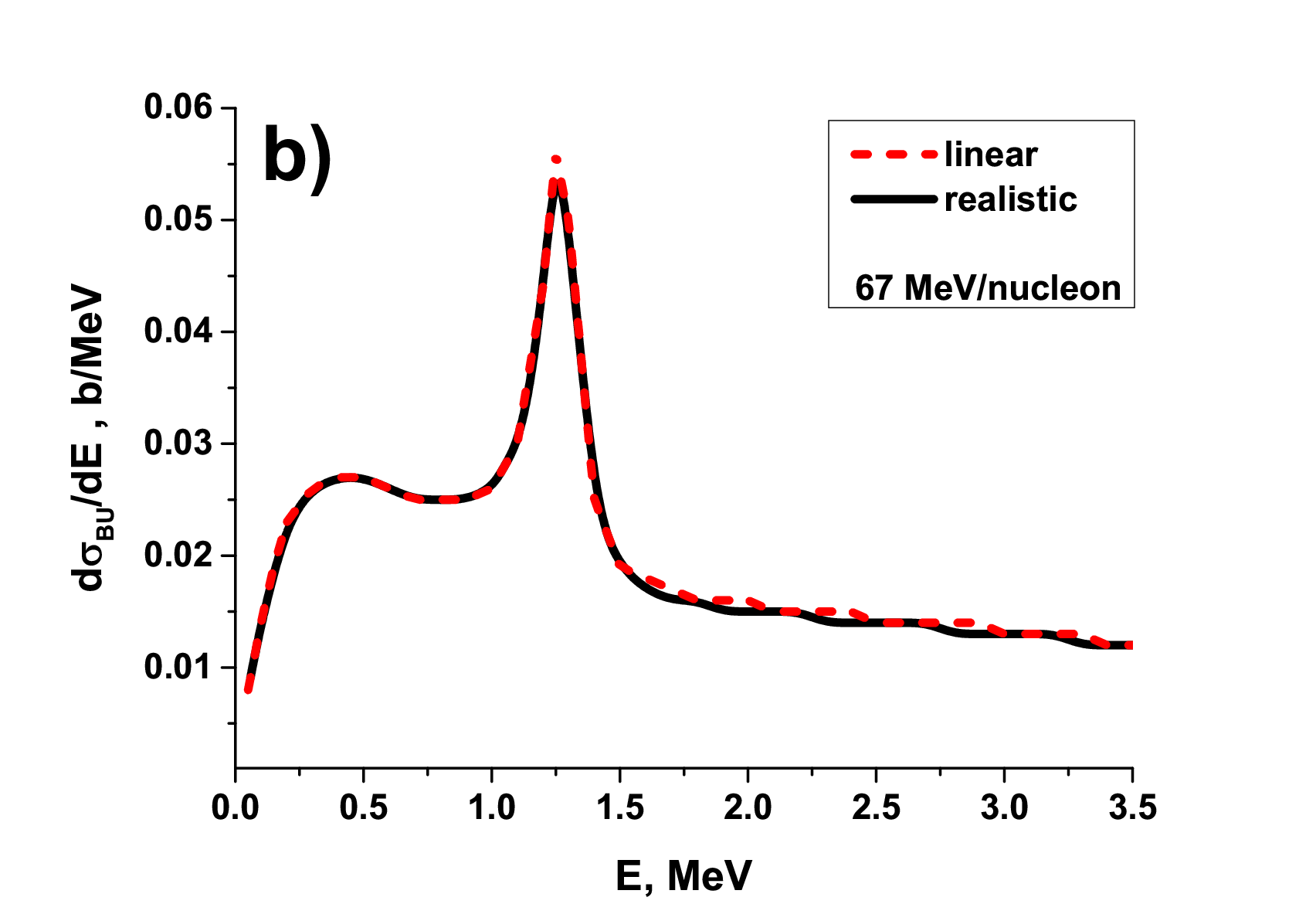}}
\end{minipage}
\caption{The comparison of breakup cross sections calculated with the time-dependent approach with linear trajectories of the projectile and with quantum-quasiclassical approach with realistic trajectories of the projectile for beam energies of 5 and 67 MeV/nucleon.}
\label{fig:4}
\end{figure}

In Fig. 2, we demonstrate the influence of the projectile curvature on the breakup cross sections. Here, the breakup cross sections calculated with the hybrid quantum-quasiclassical approach (realistic trajectories) and with time-dependent approach with linear trajectories are presented for beam energies 5 MeV/nucleon (graph a) and 67 MeV/nucleon (graph b). In both calculations two bound states ($1/2^{+}$, $1/2^{-}$) and three low-lying resonances ($5/2^{+}$, $3/2^{-}$, $3/2^{+}$) of $^{10}$Be+n system were included. The Coulomb and nuclear interactions between the projectile and the target (3) were also included. The results given in Fig.2 show that in the energy range 10 - 67 MeV/nucleon the contribution of the nonlinearity of the projectile curve does not exceed two percent. Nevertheless, at low beam energy (5 MeV/nucleon), a decrease in the ($d\sigma_{bu}(E)/dE$) cross section calculated with real trajectories in the region 0.3 MeV $\leq E \leq $ 1.1 MeV is the order of 6 - 7 $\%$ and about 3 $\%$ in E$\geq$1.7 MeV (in the region of low-lying resonances $3/2^{-}$ and $3/2^{+}$ with resonant energies $E_{3/2^{-}}$= 2.789 MeV and $E_{3/2^{+}}$= 3.367 MeV, graph (a)). Including into consideration the deformation of projectile trajectory also slightly decreases the maximum of the resonance $5/2^{+}$ at $E_{5/2^{+}}$= 1.232 MeV.
At our previous studies of the influence of the low-lying resonances into the breakup of $^{11}$Be on $^{208}$Pb, it was found that for the incident beam energies at 5 - 30 MeV/nucleon, the contribution of the $5/2^{+}$ resonance state of $^{11}$Be to the breakup cross sections is clearly visible, while at energies $\sim$70 MeV/nucleon, resonant states $3/2^{-}$ and $3/2^{+}$ make the largest contribution~\cite{valiolda2022}. It should be noted that in the case of a heavy target ($^{208}$Pb), the differences between the cross sections calculated with the linear and realistic trajectories of the projectile were about several percent in the energy range 30 - 20 MeV/nucleon, for 10 MeV/nucleon the discrepancy was 10$\%$ and reached a value of more than 20$\%$ at 5 MeV/nucleon, which exceeded the effect of nuclear interaction~\cite{valiolda2022}. Here, for a breakup on the light target ($^{12}$C), the influence of projectile trajectory curvature on the breakup cross section with decreasing collision energy is significantly less than in case of a heavy target ($^{208}$Pb) due to weaker Coulomb interaction between $^{11}$Be and $^{12}$C than between $^{11}$Be and $^{208}$Pb, as we assume.

Fig. 3 illustrates the comparison of calculated breakup cross section of $^{11}$Be on $^{12}$C with the experimental data measured by Fukuda et al. at 67 MeV/nucleon~\cite{fukuda2004} and with the results of calculations of earlier work~\cite{capel2004} performed with non-perturbative time-dependent approach~\cite{melezhik1999, melezhik2003}. Both the theoretical breakup cross sections at energy $E_{0}$ are convoluted with the instrumental energy resolution $(E_{0}^{-1/2}/0.48exp[-(E-E_{0})^{2}/0.073E_{0}]$, with $E_{0}$ and $E$ in MeV)~\cite{capel2004} to obtain a value at energy $E$ comparable to experiment. The major effect of this convolution is to resize the peaks, that are broadened and slightly shifted towards lower energies. In both computations the parameters of optical potentials and the edge parameters of integral (7) are the same except the angular basis functions as we use an angular grid with the equal numbers $N_{\theta}=N_{\varphi}=$15 of grid return $N=N_{\theta} \times N_{\varphi}$ over $\theta$ and $\varphi$, whereas in~\cite{capel2004} the authors applied angular basis with $N_{\theta}=$12,  $N_{\varphi}=$23. Our results (red full line) were obtained with including of two bound ($1/2^{+}$, $1/2^{-}$) and three resonance ($5/2^{+}$, $3/2^{-}$ and $3/2^{+}$) states of $^{11}$Be into hybrid quantum-quasiclassical approach accounting the curvilinear trajectories of the projectile. Otherwise, in calculations of~\cite{capel2004}, two bound and only one resonance $5/2^{+}$ were included in the time-dependent approach with linear trajectories of the projectile. As it is shown at Fig.3, the computation of the breakup cross section, adjusted for curvature of the projectile trajectory due to the interaction with the target and including low-lying resonances ($5/2^{+}$, $3/2^{-}$ and $3/2^{+}$), provides slightly better description of the experimental data at the region of resonance $5/2^{+}$.

\begin{figure}[htp]
\centerline{\includegraphics[width=11cm]{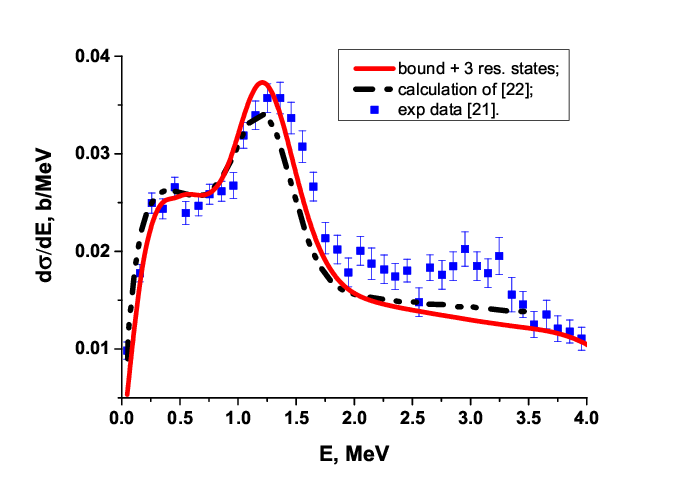}}
\caption{ The comparison of theoretical~\cite{capel2004} and experimental~\cite{fukuda2004} breakup cross sections with our results, obtained with quantum-quasiclassical approach, including two bound ($1/2^{+}$, $1/2^{-}$) and three resonance ($5/2^{+}$, $3/2^{-}$, $3/2^{+}$) states of $^{11}$Be at 67 MeV/nucleon.}
\label{fig:3}
\end{figure}

\begin{figure}[htp]
\centerline{\includegraphics[width=11cm]{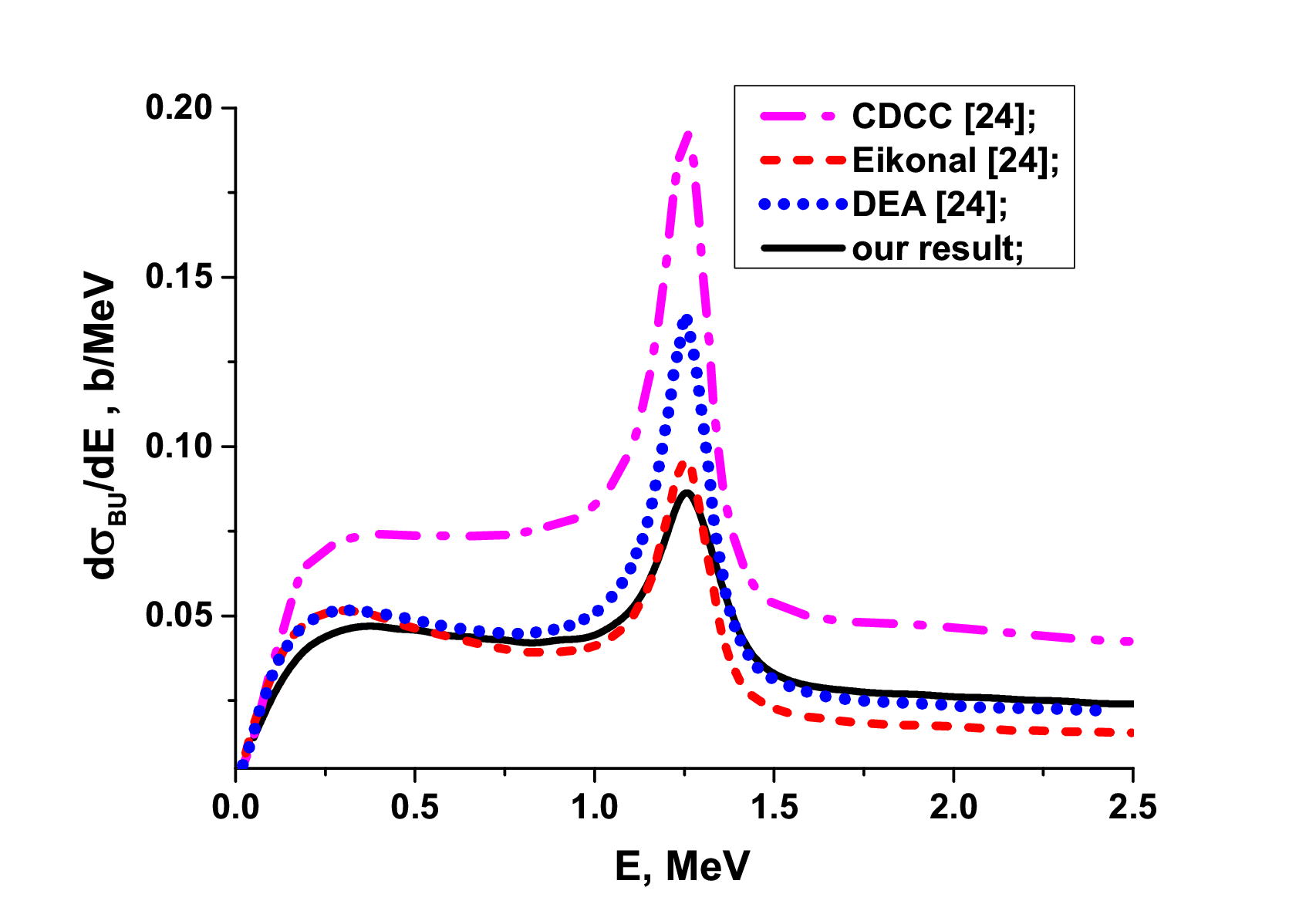}}
\caption{Breakup cross sections of $^{11}$Be on $^{12}$C at 20 MeV/nucleon calculated within quantum-quasiclassical approach in comparison with other theoretical approaches from~\cite{hebborn2018}.}
\label{fig:3}
\end{figure}

To date, there are no experimental data for low beam energies for the breakup reaction of $^{11}$Be on carbon and also on lead targets. Nevertheless, attempts of calculations at lower energies have already been carried out within the framework of CDCC, DEA and eikonal approaches~\cite{hebborn2019, hebborn2018}. However, it was concluded by Hebborn and Capel~\cite{hebborn2018} that they did not improve the description of the breakup of halo nuclei with the eikonal approaches down to 20 MeV/nucleon and "CDCC exhibits convergence issues in this range of energies". Nevertheless, in the absence of other results, we made comparisons with these calculations.    
In Fig.4, we present breakup cross sections of $^{11}$Be on $^{12}$C calculated at 20 MeV/nucleon as a function of the $^{10}$Be+n relative energy. They are plotted in comparison with the coupled-channel technique with a discretized continuum (CDCC) (a purple dotted-dashes), the eikonal model (a red dashes line) and dynamical eikonal approximation (DEA) (a blue dotted line)~\cite{hebborn2018}. Our results, highlighted with a solid line, are obtained with including two bound ($1/2^{+}$, $1/2^{-}$) and three resonant ($5/2^{+}$, $3/2^{-}$, $3/2^{+}$) states of $^{10}$Be+n system in the frame of the quantum-quasiclassical approach with curvilinear trajectories of the projectile.
Theoretical calculations from~\cite{hebborn2018} were performed with including only one resonance $5/2^{+}$ and two bound ($1/2^{+}$, $1/2^{-}$) states of the core-neutron system. As it is shown in Fig.4, the CDCC result~\cite{hebborn2018} exceeds our cross sections by approximately 30$\%$ over the whole energy range. Otherwise, DEA result exceeds essentially values of breakup cross sections in the peak (corresponding to the resonance position of $5/2^{+}$), calculated with eikonal approximation~\cite{hebborn2018} and our quantum-quasiclassical approach. Our results agree well with the eikonal approximation~\cite{hebborn2018} at $^{10}$Be+n relative energies up to 1.3 MeV and with DEA at energies above 1.4 MeV. Note, that transition in our approach to the straight-line trajectory leads to visible exceeding of our cross sections below 1.5 MeV (see Fig.2(a)). In the case of the eikonal approximations~\cite{hebborn2018}, the corrections to the straight-line trajectories goes in the same direction as what we found using our quasi-classical approach.

In Fig. 5 we present the breakup cross sections of $^{11}$Be on the $^{12}$C target calculated for two bound states ($1/2^{+}$, $1/2^{-}$) at broad range 5 - 67 MeV/nucleon of beam energies. We took into account the influence of the resonant states ($5/2^{+}$, $3/2^{-}$, $3/2^{+}$) and the effect of the deformation of the projectile trajectory, as well as the transfer of energy from the target to the projectile and vice versa during the breakup process~\cite{valiolda2022}.

\begin{figure}[htp]
\centerline{\includegraphics[width=11cm]{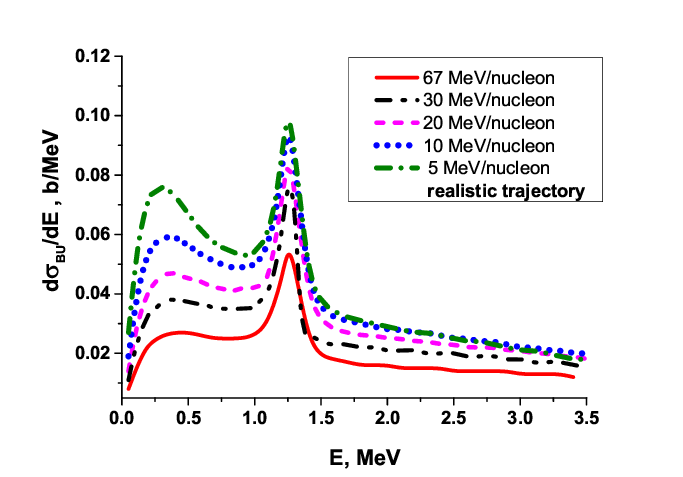}}
\caption{The breakup cross section $d\sigma_{bu}(E)/dE$ calculated as a function of beam energy and the energy between $^{10}$Be and neutron, including two bound ($1/2^{+}$, $1/2^{-}$) states and three resonances ($5/2^{+}$, $3/2^{-}$, $3/2^{+}$) of $^{11}$Be.}
\label{fig:3}
\end{figure}

\section{Investigation of spectral structure of $^{11}$Be in breakup reactions} 

Let us note two features in the dynamics of the breakup cross section of $^{11}$Be with decreasing the colliding energy (see Fig.5): a dramatic increase of the $5/2^{+}$ resonance at 1.25 MeV and the transformation of the plateau in the 0.3 - 1.0 MeV region into a broad peak at ~0.3 MeV as the colliding energy decreases to 5 MeV/nucleon. To clarify the physical nature of these two peaks, we compared the calculated cross section $^{11}$Be+$^{12}$C $\rightarrow ^{10}$Be+n+$^{12}$C at 5 MeV/nucleon with the cross section $^{11}$Be+$^{208}$Pb $\rightarrow ^{10}$Be+n+$^{208}$Pb calculated in our previous work~\cite{valiolda2022} at the same energy (see Fig.6).

\begin{figure}[htp]
\centerline{\includegraphics[width=11cm]{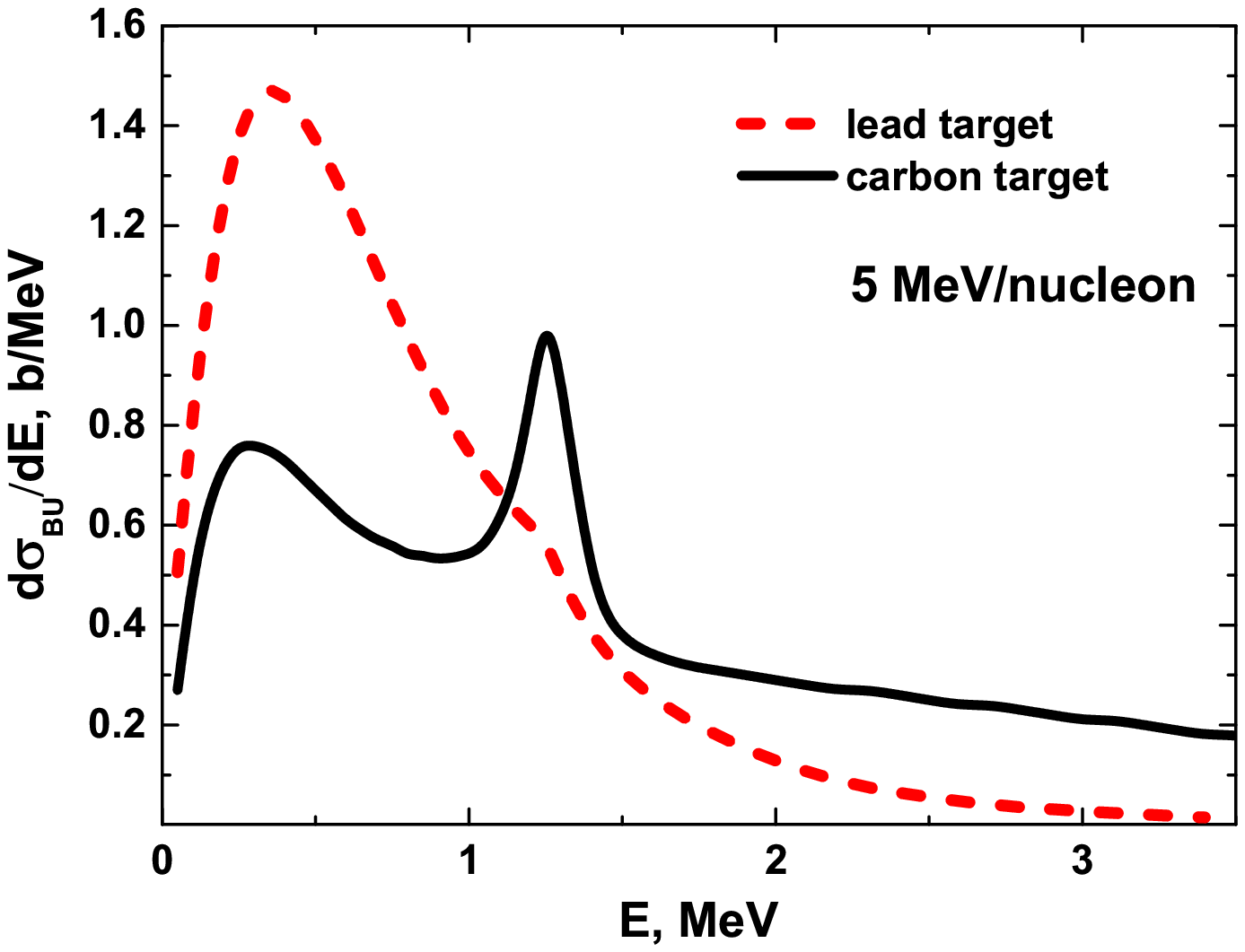}}
\caption{Breakup cross sections of $^{11}$Be impinging on light ($^{12}$C) (multiplied by 10) and heavy ($^{208}$Pb) targets calculated within quantum-quasiclassical approach at 5 MeV/nucleon.}
\label{fig:3}
\end{figure}

\noindent In the latter case, the peak at 1.25 MeV is only slightly noticeable, although the breakup cross section on the heavy target is significantly higher than the cross section on the light target at the region below 2 MeV. However, the fact that the positions of the 1.25 MeV and 0.3 MeV peaks do not depend on the type of the target leads to the natural conclusion that they are related to the spectral structure of $^{11}$Be nucleus. Since the quantum-quasiclassical computational scheme allows one to manipulate the spectrum of $^{11}$Be, i.e. by removing the excited state $1/2^{-}$ from the corresponding partial interaction n-$^{10}$Be and varying the binding energy of the ground state $1/2^{+}$~\cite{valiolda2022}, this makes it possible to study the influence of these bound states on the peak in the cross section about 0.3 MeV. The results of this investigation are given in Fig.7. Here it can be seen that the elimination of the excited state $1/2^{-}$ from the n-$^{10}$Be interaction leads to an increase in the peak at 0.3 MeV, since the breakup from the ground state $1/2^{+}$ goes directly into the continuous spectrum without intermediate population of the excited state $1/2^{-}$. A further change in the interaction, leading to an increase of the binding energy of the ground state of $^{11}$Be from $E_{1/2^{+}}$= -0.503 MeV, leads to the transformation of the peak and to a plateau of the cross section at $E_{1/2^{+}}$= -1 MeV and then to the elimination of the plateau at $E_{1/2^{+}}$ $\leq$ -2 MeV. The influence of the position of the ground state energy level on the resonance amplitude is also noticeable for $5/2^{+}$ resonance at 1.25 MeV.

\begin{figure}[htp]
\centerline{\includegraphics[width=11cm]{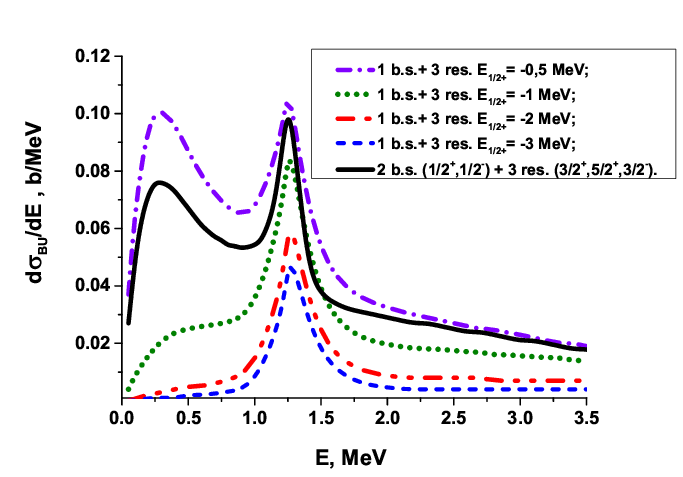}}
\caption{Breakup cross sections of $^{11}$Be on $^{12}$C with different spectral structure of $^{11}$Be nucleus calculated at 5 MeV/nucleon.}
\label{fig:3}
\end{figure}

\begin{figure}[htp]
\begin{minipage}[h]{0.47\linewidth}
\center{\includegraphics[width=1\linewidth]{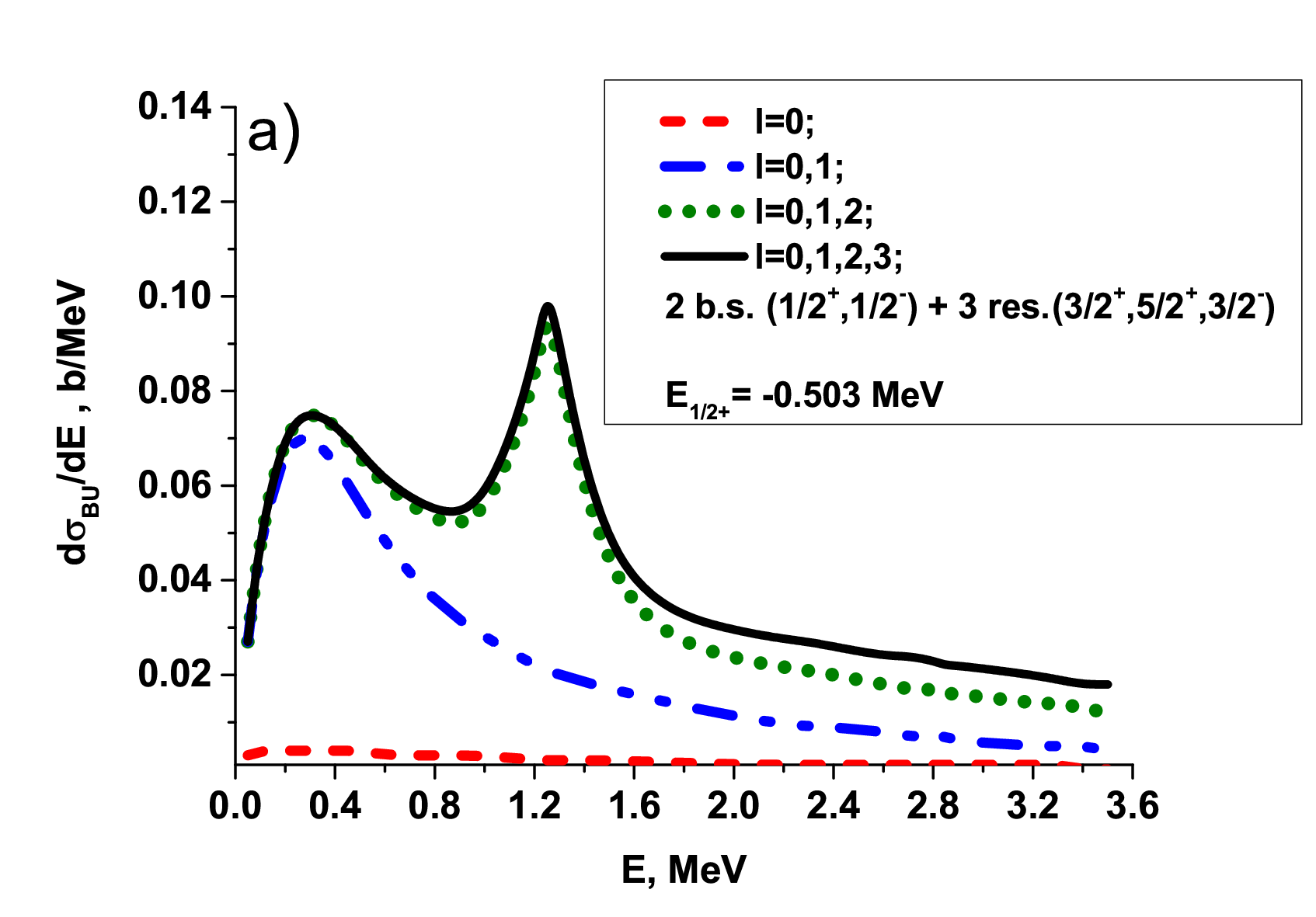}} \\
\end{minipage}
\hfill
\begin{minipage}[h]{0.47\linewidth}
\center{\includegraphics[width=1\linewidth]{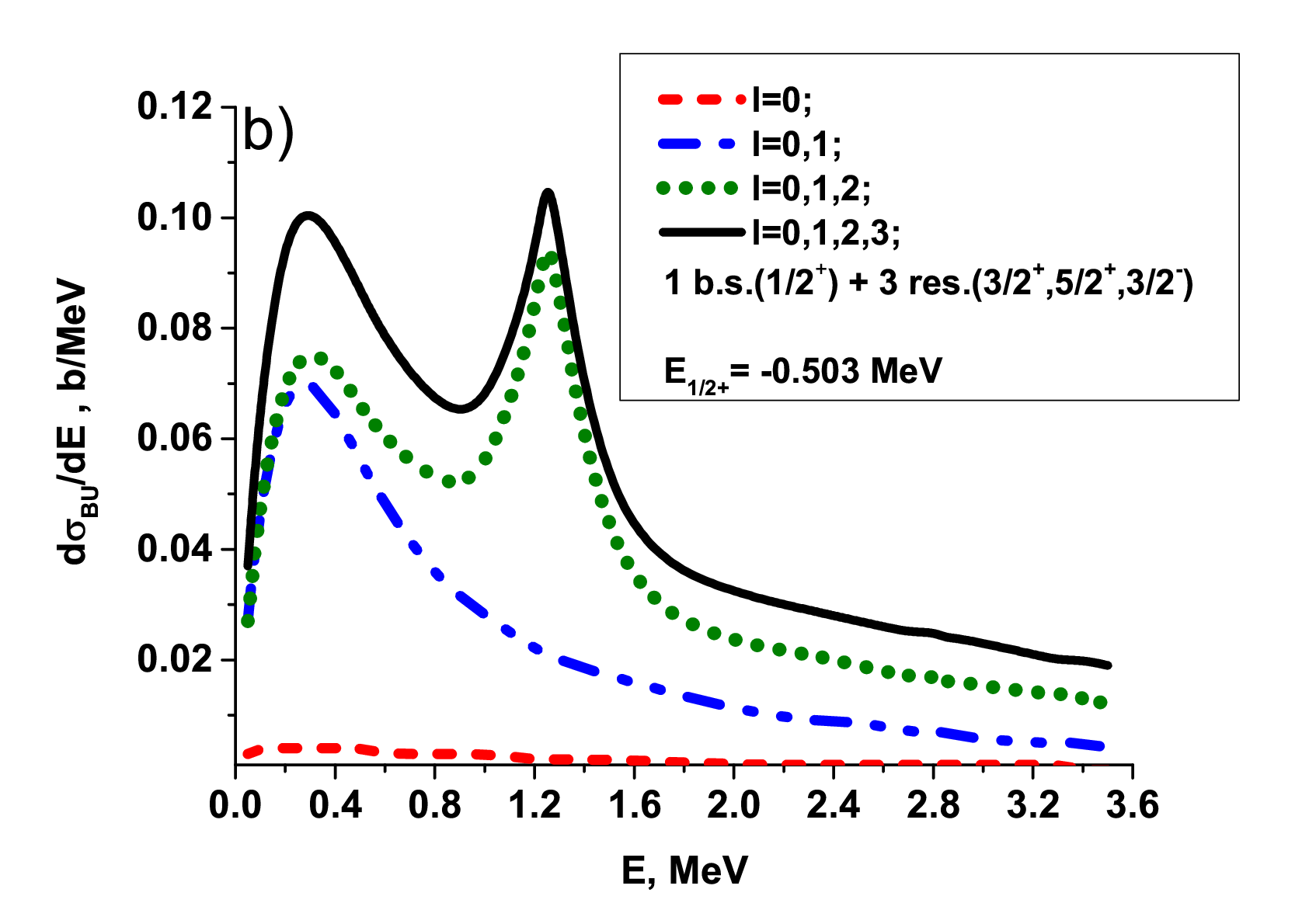}} \\
\end{minipage}
\vfill
\begin{minipage}[h]{0.47\linewidth}
\center{\includegraphics[width=1\linewidth]{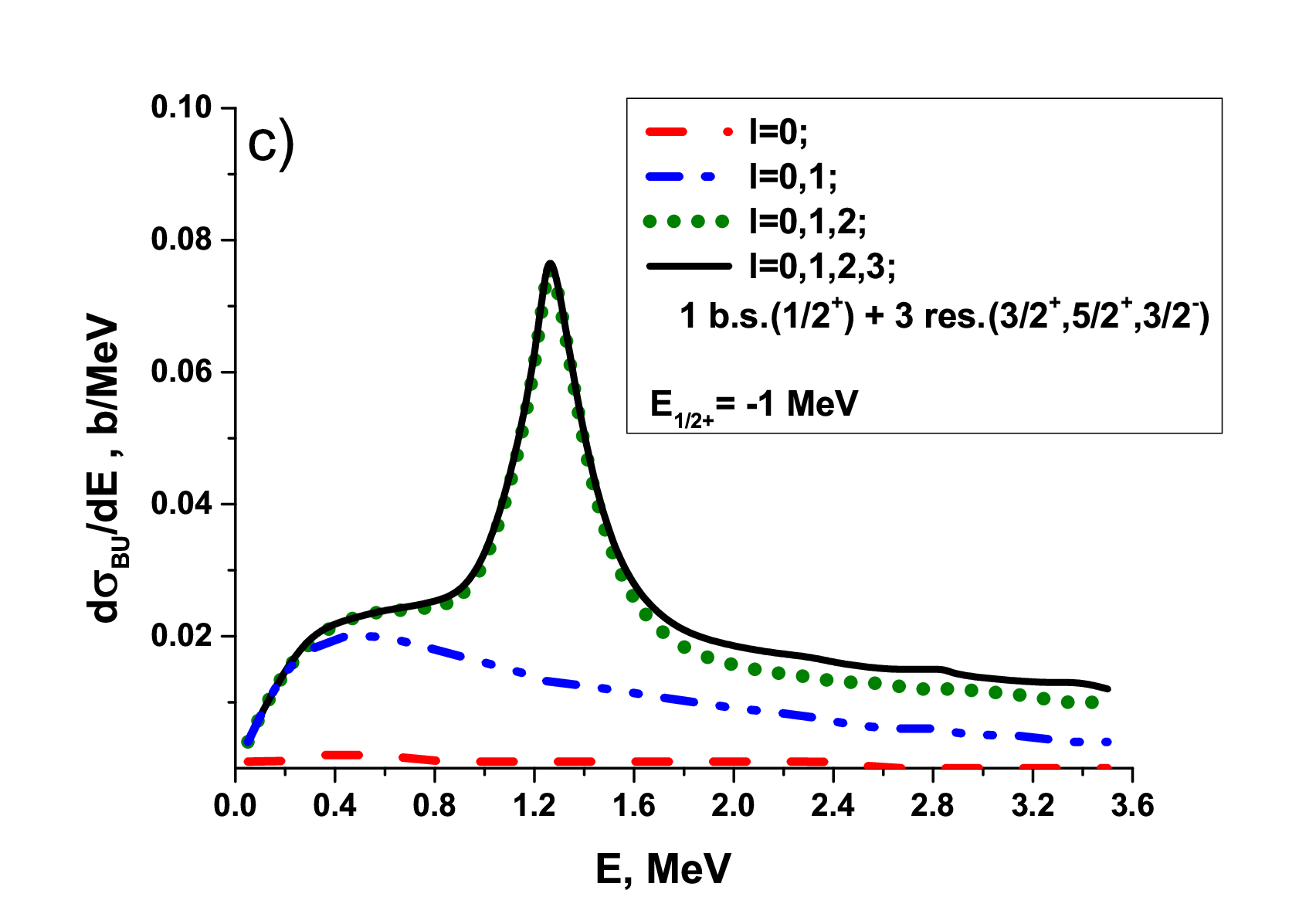}} \\
\end{minipage}
\hfill
\begin{minipage}[h]{0.47\linewidth}
\center{\includegraphics[width=1\linewidth]{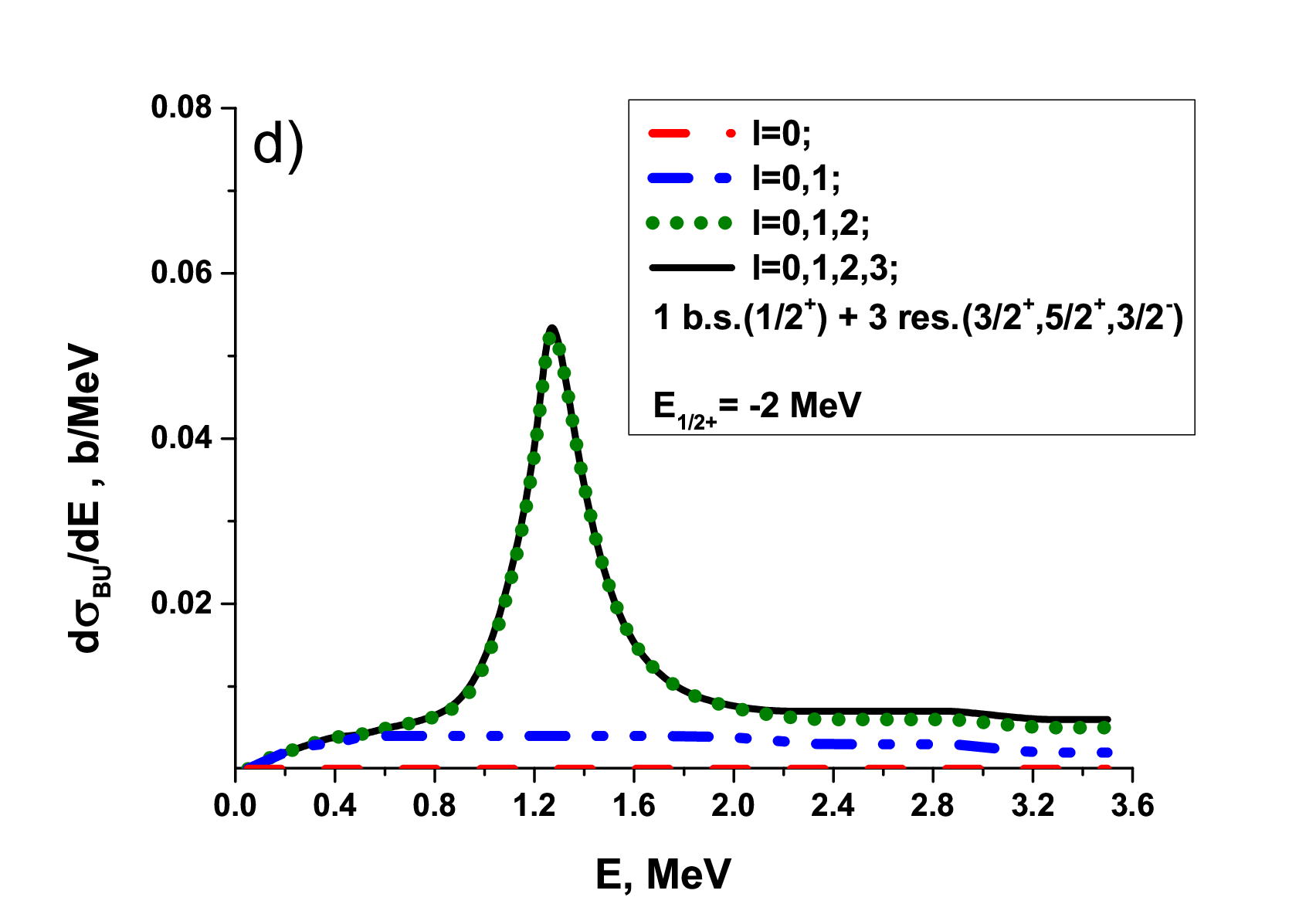}} \\
\end{minipage}
\caption{The breakup cross section $d\sigma_{bu}(E)/dE$ of $^{11}$Be at 5 MeV/nucleon calculated with increasing number of partial waves $lm$ in the sum of Eq.(7). The calculations were performed with the complete number of $m$ for every $l$-shell. Figure (a) shows the calculation results for the case of two bound states 1/2$^+$ and 1/2$^-$ in the $^{11}$Be Hamiltonian Eq.(2). The case of one bound state 1/2$^{+}$ in the $^{11}$Be Hamiltonian is shown in Figure (b). Figure (c) and (d) illustrates the calculation of breakup cross sections for deeper energy levels of 1/2$^+$ as $E_{1/2^{+}}$= -1 MeV (c) and $E_{1/2^{+}}$= -2 MeV (d) with including one bound (1/2$^+$) and 3 resonance states.}
\label{ris:experimentalcorrelationsignals}
\end{figure}

The results in Fig.8 illustrate the contribution of different partial waves in the continuum to the breakup cross sections at 5 MeV/nucleon calculated according to Eq.(7). We separated this study into two parts. In Fig.8a we demonstrate the convergence of the breakup cross section over partial waves in the continuum, calculated with a successive increase in the number of partial waves $l$ in Eq.(7) for the case of two bound states 1/2$^+$ and 1/2$^-$ in the $^{11}$Be Hamiltonian Eq.(2). In Fig.8b the case of one bound state 1/2$^+$ in the $^{11}$Be Hamiltonian is illustrated. We see that the p-wave gives the main contribution to the peak at low energy below 0.5 MeV, and the d-wave mainly contributes in the 5/2$^+$ resonance and above (see Fig.8a).
The contribution of the s-wave is insignificant in the entire energy range under considered. Influence of f-wave becomes significant only above 1.6 MeV.

The excluding of the excited state 1/2$^-$ from the $^{11}$Be Hamiltonian practically does not change the contribution of the s-wave in the entire energy range under consideration and very slightly decreases the contributions of the p- and d-waves (see Fig.8b). Thus, a comparison of the two calculations given in Fig.8a and Fig.8b shows that the emitted neutron ends up in the p- and d-waves of the continuum spectrum due to direct transitions 1/2$^+ \rightarrow kp$ and 1/2$^+ \rightarrow kd$ from the ground state of the $^{11}$Be nucleus. The transitions 1/2$^+ \rightarrow$ 1/2$^-\rightarrow kp$ and 1/2$^+\rightarrow$ 1/2$^- \rightarrow kd$ through the excited state 1/2$^-$ of $^{11}$Be give an insignificant contribution to the population of the states $kp$ and $kd$. In this regard, the mechanism of population of the $kf$ state by a neutron is of interest due to visible essential effect of the excited state 1/2$^-$ of $^{11}$Be on the population of the f-wave of the continuum at the energies below 1.6 MeV. Indeed, from Fig.8b we see that the direct transition 1/2$^+\rightarrow kf$ gives near the peak at 0.3 MeV about 35\% of the contribution from the transition 1/2$^+\rightarrow kp$ to the total breakup section (see Fig.8b). However, the including of the excited state 1/2$^-$ in the $^{11}$Be Hamiltonian  leads practically to complete suppression of the contribution of the f-wave to the total breakup cross section below 1.6 MeV (see Fig.8a). We interpret this small contribution of the f-wave in the total breakup cross section as a mutual compensation of the direct transition 1/2$^+\rightarrow kf$ and the transition 1/2$^+\rightarrow$ 1/2$^- \rightarrow kf$ through the intermediate excited state 1/2$^-$ of $^{11}$Be.

We explain the anomalously large contribution of the $kf$ states of the continuum to the breakup cross section (7) (see Fig.8 (b)) by the anomalously long tail of the weakly-bound state $s$1/2$^+$ with energy $E_{1/2^{+}}$= -0.5 MeV for the halo nucleus $^{11}$Be. To confirm this we performed calculations for cases when the potential $V_{0}(r)$ has deeper energy levels $E_{1/2^{+}}$ with $E_{1/2^{+}}$= -1 MeV and -2 MeV, and the potential $V_{1}(r)$ is zero. In these cases, the halo effect disappears (i.e., the tail of the wave function of the weakly-bound state $s$1/2$^+$ is pulled under the potential barrier): the anomalously large contribution of the wave $kf$ to the breakup cross section disappears with an increase in the nuclear binding energy (see Fig. 8 (c,d)).

The study shows that the broad peak in the breakup cross section of $^{11}$Be on $^{12}$C near 0.3 MeV at 5 MeV/nucleon beam energy is caused by a weakly-bound ground state $1/2^{+}$ of $^{11}$Be with the energy $E_{1/2^{+}}$= -0.503 MeV due to the $1/2^{+}$ $\rightarrow kp$ to the p-state of the continuum and the presence of the excited state $1/2^{-}$ decreases its maximum by an order of 35\% due to the destructive interference between the transitions $1/2^{+}$ $\rightarrow kf$ and $1/2^{+}$ $\rightarrow$ $1/2^{-}$ $\rightarrow kf$.

\section{Conclusion} 

In this work the theoretical study is performed to the breakup of the $^{11}$Be halo nuclei on a light target ($^{12}$C) from intermediate (67 MeV/nucleon) to low (5 - 30 MeV/nucleon) energies within quantum-quasiclassical approach, in which the three dimensional time-dependent Schr\"odinger  equation for halo nucleon was integrated simultaneously with the classical Hamiltonian equations describing relative projectile-target dynamics. In this hybrid quantum-quasiclassical approach the time-dependent Schr\"odinger equation is integrated numerically with a technique developed to study the Coulomb breakup of halo nuclei~\cite{melezhik1999, melezhik2003}.

In the frame of the quantum-quasiclassical approach we have calculated the breakup cross sections of $^{11}$Be on a carbon target at energies 5 - 67 MeV/nucleon including Coulomb and nuclear interaction between projectile and target. The performed research demonstrates that the model with the straight-line projectile trajectories provides  a satisfactory accuracy in calculating the breakup cross sections of $^{11}$Be with decreasing the beam energy down to 5 - 10 MeV/nucleon. The uniqueness of our calculations lies in the inclusion of low-lying resonances ($5/2^{+}$, $3/2^{-}$, $3/2^{+}$) in the breakup cross section of the $^{11}$Be nucleus~\cite{valiolda2022, valipepan2022, valejpfm2022, valejpfm2023}. We also  show that the analysis with our quantum-quasiclassical approach allows to investigate spectral properties of $^{11}$Be nuclei from the breakup reactions. Our results describe well the existing experimental data of Fukuda et al.~\cite{fukuda2004} at 67 MeV/nucleon and are in comparative agreement with other existing calculations performed with alternative  theoretical models at 67 MeV/nucleon~\cite{capel2004} and 20 MeV/nucleon~\cite{hebborn2018}.

Summarizing, the results obtained by quantum-quasiclassical  approach can potentially be useful in further investigations of breakup reactions at low energies. In particular the region around 20 - 10 MeV/nucleon is of great interest, since this is the energy range of HIE-ISOLD at CERN and the future ReA12 at MSU, it has hardly been investigated theoretically so far.

\section*{Acknowledgments}

This research is funded by the Science Committee of the Ministry of Science and higher education of the Republic of Kazakhstan (Grant No. AP13067742). Additionally, authors note that the convergence of computational scheme and its optimization was investigated with the algorithm developed with support of the Russian Science Foundation under the Grant No. 20-11-20257.

\end{document}